# Classification of diffraction patterns using a convolutional neural network in single particle imaging experiments performed at X-ray free-electron lasers

Dameli Assalauova[1,+], Alexandr Ignatenko[1,+], Fabian Isensee[2,3,+], Sergey Bobkov[4], Darya Trofimova[2,3], and Ivan A. Vartanyants[1,5,*]

[1]*Deutsches Elektronen-Synchrotron DESY, Notkestrasse 85, 22607 Hamburg, Germany;*

[2]*Applied Computer Vision Lab, Helmholtz Imaging, German Cancer Research Center DKFZ, Im Neuenheimer Feld 280, 69120 Heidelberg, Germany;*

[3]*Division of Medical Image Computing, German Cancer Research Center DKFZ, Im Neuenheimer Feld 280, 69120 Heidelberg, Germany;*

[4]*National Research Center "Kurchatov Institute", Akademika Kurchatova pl. 1, 123182 Moscow, Russian Federation;*

[5]*National Research Nuclear University MEPhI, Kashirshkoe sh. 31, 115409 Moscow, Russia*

## Abstract

Single particle imaging (SPI) at X-ray free electron lasers (XFELs) is particularly well suited to determine the 3D structure of particles in their native environment. For a successful reconstruction, diffraction patterns originating from a single hit must be isolated from a large number of acquired patterns. We propose to formulate this task as an image classification problem and solve it using convolutional neural network (CNN) architectures. Two CNN configurations are developed: one that maximises the F1-score and one that emphasises high recall. We also combine the CNNs with expectation maximization (EM) selection as well as size filtering. We observed that our CNN selections have lower contrast in power spectral density functions relative to the EM selection, used in our previous work. However, the reconstruction of our CNN-based selections gives similar results. Introducing CNNs into SPI experiments allows streamlining the reconstruction pipeline, enables researchers to classify patterns on the fly, and, as a consequence, enables them to tightly control the duration of their experiments. We think that bringing non-standard artificial intelligence (AI) based solutions in a well-described SPI analysis workflow may be beneficial for the future development of the SPI experiments.

[+]These authors contributed equally to this work

[*] Corresponding author: Ivan.Vartaniants@desy.de

**I. Introduction**

Artificial Intelligence (AI) and Machine Learning (ML) methods are rapidly becoming an important tool in physics research. We witness an increased interest in these approaches, especially, during the last years. This is also related to a large amount of data collected nowadays in the experiments not only in particle physics, but also in astronomy or X-ray physics. For example, in a few days of measurements at the megahertz European X-ray Free-Electron Laser (EuXFEL) at one beamline petabytes of data may be easily collected. Machine learning approaches can help us to use effectively the most out of this huge amount of data.

One of the flagship experiments at XFEL is the so-called Single Particle Imaging (SPI). In these experiments single biological particles such as viruses or protein complexes are injected in the intense femtosecond XFEL beam in their native environment and diffraction patterns are collected before particles are disintegrated due to Coulomb explosion (Neutze *et al.*, 2000). By collecting a sufficient number of diffraction patterns originating from reproducible biological samples at different orientations the full three-dimensional (3D) diffracted intensity may be constructed and then, applying phase retrieval techniques, a high resolution image of the biological sample may be obtained (Gaffney & Chapman, 2007). Being well defined, the task of obtaining high resolution images of single biological particles at XFEL is still far from being solved. In order to determine the best strategies to push SPI to higher resolution the SPI consortium was formed at LCLS (Stanford, USA) (Aquila *et al.*, 2015).

In the framework of this consortium several strategies for the data analysis were developed. Typical SPI data analysis comprises several sequential steps from the raw detector images to 3D reconstructed particle structure (see Fig. 1). This workflow consists of the following steps: initial pre-processing of diffraction patterns, particle size filtering, single hit diffraction patterns classification, orientation determination and obtaining of the 3D intensity map of the particle, and, finally, phase retrieval and reconstruction of the 3D electron density of the biological sample (Gaffney & Chapman, 2007; Rose *et al.*, 2018; Assalauova *et al.*, 2020). An important step in this data processing pipeline is single hit classification. Only diffraction patterns that contain the scattering signal of a single particle are of interest for the further analysis. In our previous work (Assalauova *et al.*, 2020) this step was addressed with the expectation-maximization (EM) algorithm, first developed in cryogenic electron microscopy (cryo-EM) (Dempster *et al.*, 1977). The EM algorithm allows for unsupervised clustering of data when neither initial data assignments to clusters nor cluster parameters are known. In the end the clusters that correspond to single hits of an investigated particle are selected manually by an expert.

The step of single hit classification may be significantly improved by application of Machine Learning approaches. Convolutional neural networks (CNNs) are now the de-facto state-of-the-art in image classification (Krizhevsky *et al.*, 2012), object detection (Szegedy *et al.*, 2013), and image segmentation (Long *et al.*, 2015). CNN-based solutions have been recently successfully applied to the classification of diffraction patterns in tomography experiments at synchrotron sources (Yang *et al.*, 2020), coherent diffraction imaging experiments at synchrotron facilities (Wu, Yoo *et al.*, 2021; Wu, Juhas *et al.*, 2021), and at XFELs (Shi *et al.*, 2019; Zimmermann *et al.*, 2019). As we showed in our previous work (Ignatenko *et al.*, 2021), a CNN-based solution can be successfully applied to the single hit diffraction patterns classification step (Fig. 1, blue arrows).

In this work, we further develop this approach (Fig. 1, red arrows). By classifying single hits first, computationally intensive steps of the pipeline, such as size filtering and EM-based selection must only be performed on a fraction of the initially collected patterns, saving substantial computational resources. In addition, the proposed scheme allows the classification of newly collected patterns independently, without the need to recompute from the beginning (as would be needed by pure EM-based selection). This is particularly useful, as experimentalists have the possibility to plan the experiment as it goes and stop it whenever a sufficient amount of single hits has been collected and by that saving the precious beamtime at XFEL facility.

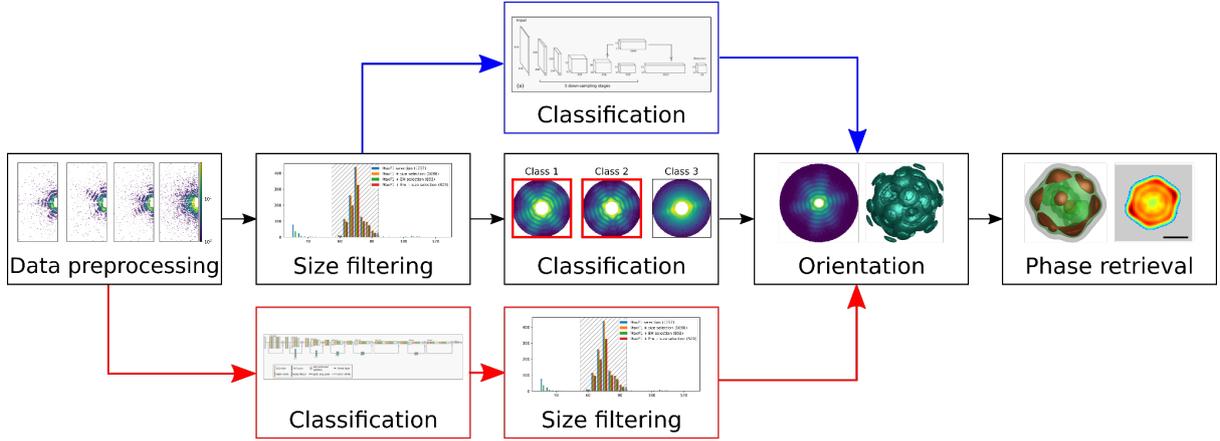

**Figure 1.** SPI workflow. Black arrows indicate the typical steps in SPI data analysis Assalauova et al., 2020). Blue arrows show the implementation of CNN-based single hit diffraction patterns classification (Ignatenko et al., 2021). Red arrows show the modified workflow for CNN-based classification prior to the particle size filtering step (this work).

## II. SPI experiments and data analysis

The SPI experiment (Fig. 2(a)) was performed at the Atomic Molecular Optics (AMO) instrument (Ferguson *et al.*, 2015; Osipov *et al.*, 2018) at the Linac Coherent Light Source (LCLS) at SLAC National Accelerator Laboratory in the frame of the SPI initiative (Aquila *et al.*, 2015). Samples of PR772 bacteriophage (Reddy *et al.*, 2017; Li *et al.*, 2020) were aerosolized using a gas dynamic virtual nozzle in a helium environment (Nazari *et al.*, 2020). The particles were injected into the sample chamber using an aerodynamic lens injector (Hantke *et al.*, 2014; Benner *et al.*, 2008). The particle stream intersected the pulsed and focused XFEL beam. The XFEL had a repetition rate of 120 Hz, an average pulse energy of ∼2 mJ, a focus size of ∼1.5 μm, and a photon energy of 1.7 keV (wavelength 0.729 nm). Diffraction patterns were recorded by a pn-type charge coupled device (pnCCD) detector (Strüder *et al.*, 2010) mounted at 0.130 m distance from the interaction region. The detector consisted of two panels. The size of each panel was 512 by 1024 pixels with a pixel size of 75 × 75 μm$^2$. The scattering signal was only recorded by one (upper) of the two detector panels (the lower one was not operational during the experiment due to an electronic fault).

The total number of diffraction patterns collected during the experiment was $1.2 \times 10^7$ (data set $D_0$ in Table I) (the experimental data are available in Ref. (Li *et al.*, 2020)). Out of those images, only a small fraction contained any scattering patterns. Hit finding was performed using the software 'psocake' in the 'psana' framework (Damiani *et al.*, 2016). As a result, 191,183 diffraction patterns (data set $D$ in Table 1) were selected as hits from the initial set of experimental data (Li *et al.*, 2020; Assalauova *et al.*, 2020). Manual selection of single hit diffraction patterns was performed (data set $D_M$ in Table 1) and resulted in 1,393 single hit diffraction patterns (see (Li *et al.*, 2020)). This selection was used as a ground truth for CNN training in this work.

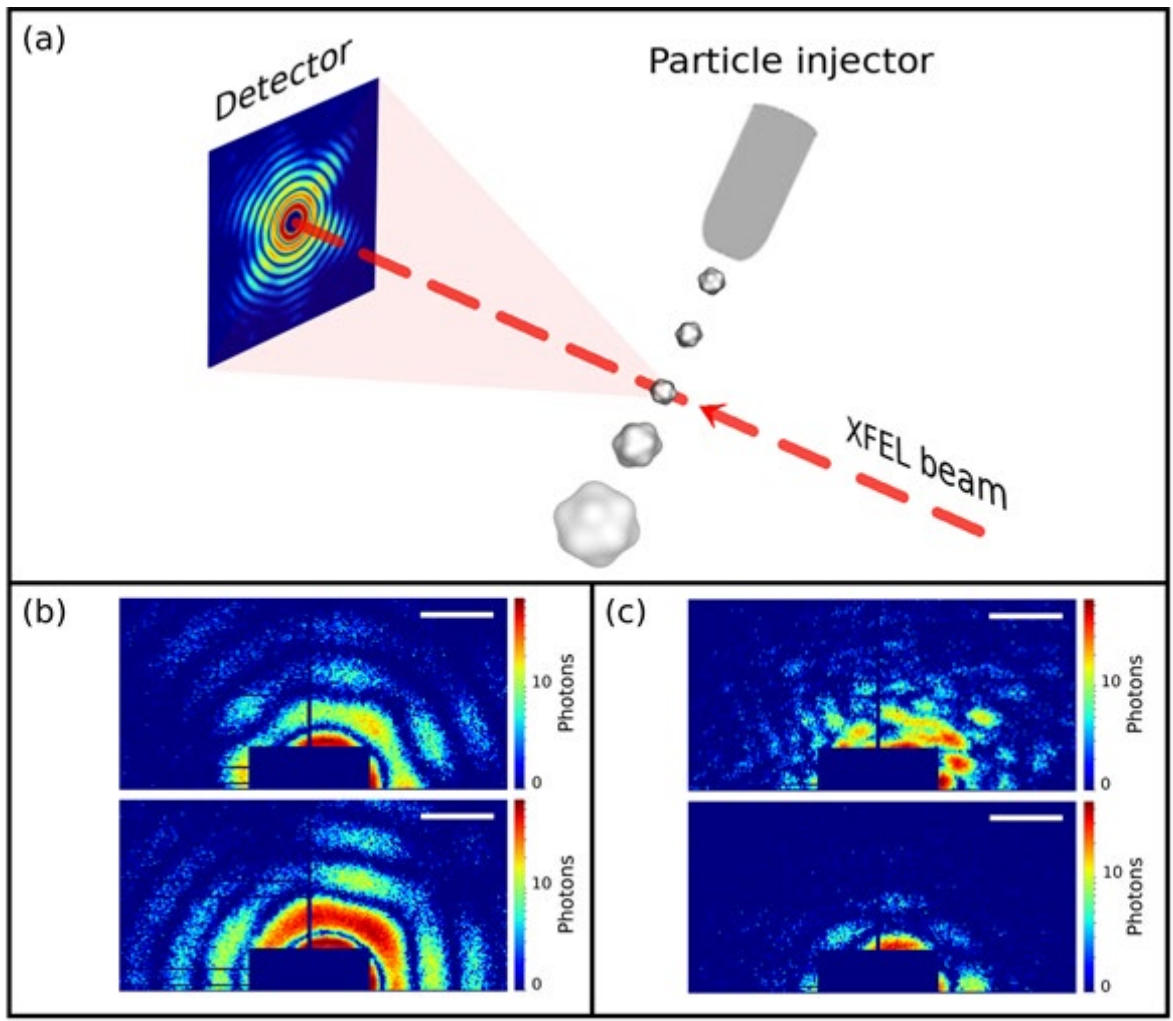

**Figure 2.** (a) Schematic representation of an SPI experiment. The incoming XFEL beam interacts with the virus injected by the particle injector. The particle is destroyed afterwards due to Coulomb explosion. X-ray radiation from the non-destroyed virus is scattered to the detector positioned in the far-field. (b) Examples of single hits. (c) Examples of non-single hits. Diffraction patterns in (b, c) are shown in logarithmic scale; the area of the size 192 x 96 pixels of the whole diffraction pattern is shown. The scale bar in (b, c) corresponds to 0.2 nm$^{-1}$.

**Table 1.** Number of diffraction patterns obtained at different SPI analysis steps.

| Data set | Number of diffraction patterns | |
|---|---|---|
| Initial data set, $D_0$ | $1.2 * 10^7$ | |
| Hit finding procedure, $D$ | 191,183 | |
| Manual selection of single hits, $D_M$ | 1,393 | |
| Selection by expectation-maximization (EM) algorithm, $D_{EM}$ | 1,085 | |
| | **Single hits** | **Non-single hits** |
| Training and validation data set, $D_{tr}$ | 100 | 19,900 |
| Test data set, $D_{test}$ | 1,293 | 169,890 |

## III. Methods

### 1. CNN architecture

The network architecture used in this work is shown in Fig. 3. It is inspired by the pre-activation ResNet-18 (He *et al.*, 2016) and was selected based on initial experiments on the training data set. The network processes patches of size 192 x 96 and is initialized with 16 convolutional filters. The number of filters is doubled with each downsampling up to a maximum of 256. Downsampling is implemented as strided convolution. We use leaky ReLU activation functions (Xu *et al.*, 2015) and standard batch normalization (Ioffe & Szegedy, 2015). The final feature map has a size of 6x6 which is aggregated through global average pooling into a vector which is then processed by a linear layer to distinguish single and non-single hits.

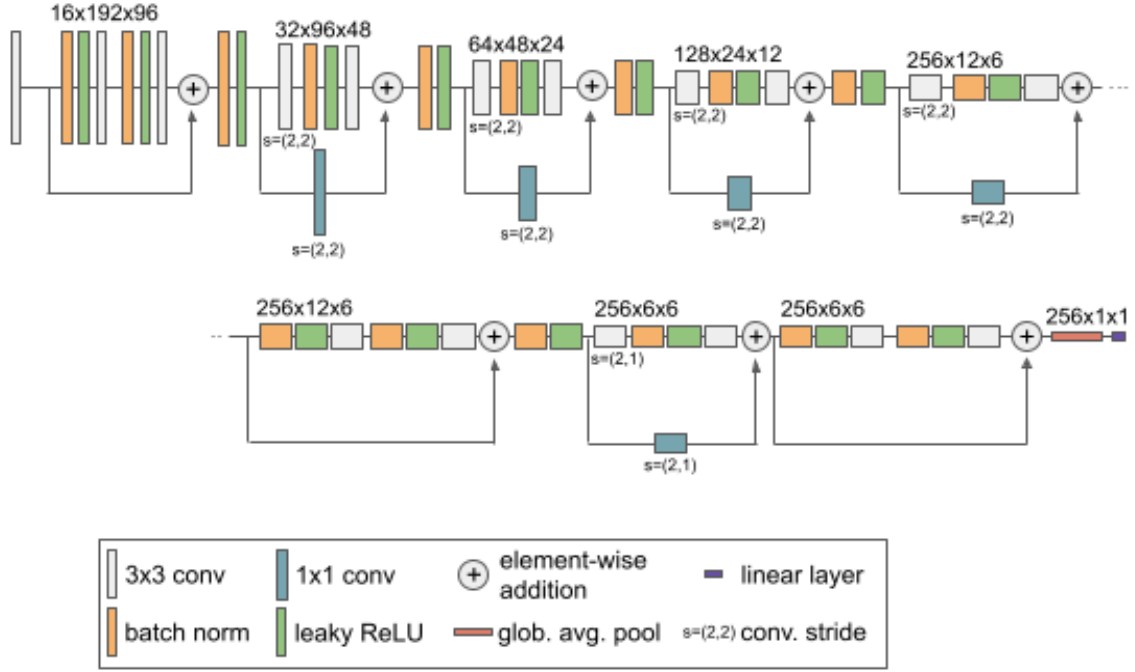

**Figure 3.** Network architecture. We use a pre-activation ResNet-inspired architecture. It takes patches of size 192 x 96 as input and processes them in a sequence of 8 preactivation residual blocks. Downsampling is implemented via strided convolution. The architecture is initialized with 16 filters and doubles the number of filters with each downsampling operation up to a maximum of 256. Global average pooling reduces the final feature representation (shape 6x6) to a vector that is then used by the classification layer to distinguish single from non-single hits. The size of the feature representations is indicated above each residual block. 16x192x96 hereby denotes 16 convolutional filters with a feature representation of size 192 x 96.

2. **CNN evaluation metrics**

As evaluation metrics we used precision, recall, and $F1$-score. These values are defined through true positive (TP), false positive (FP) and false negative (FN) predictions. The definition of the evaluation metrics is as follows

$$precision = \frac{TP}{TP + FP}, \quad (1)$$

$$recall = \frac{TP}{TP + FN},  \quad (2)$$

$$F1 = 2\frac{precision * recall}{precision + recall} = \frac{2*TP}{2*TP + FP + FN}. \quad (3)$$

Due to the pronounced class imbalance in our data set (small number of single hits in comparison to a large number of non-single hits) we mainly use the F1-score for evaluating our models. In addition, we report the number of single hits.

### 3. Training, validation and test procedure in CNN classification

We use a training data set that is representative of the modified workflow introduced in section II, where the experimentalist identifies a limited number of single hits at the beginning of the experiment. Based on the annotation effort that would be required, we chose to use 100 single hits and a number of non-single hits that corresponds to the number of images the experimentalist would have seen until the required number of single hits was collected (see Table 1). Based on the class ratio of the data set used here (approximately 1:200) our training set ($D_{tr}$) consists of 100 single and 19,900 non-single hits. All hits were sampled randomly without replacement. We used the manual selection $D_M$ as a ground truth.

To prepare our data for the CNN all diffraction patterns were cropped to an area of size 192 x 96 pixels (see Supporting Information Fig. S1 and Fig. 2(a,b)). All images were normalized by subtraction of the training data set (20,000 data) mean value (μ = 0.342) and divided by the standard deviation of the same data set (σ = 2.336).

During method development, our models were trained and validated through stratified five-fold cross-validation on the set of 20,000 training examples. We report final results on the test set ($D_{test}$) consisting of the 171,183 remaining patterns (1,293 single and 169,890 non-single hits) (see Supporting Information Section S3.3)

We train the network with stochastic gradient descent using the Adam optimizer (Kingma & Ba, 2014), a minibatch size of 64 and an initial learning rate of $10^{-4}$. The standard cross-entropy loss function is used. Samples within minibatches are sampled randomly with replacement. We modify the sampling probabilities such that on average 2% of the presented samples are single hits. We define an epoch as 50 training iterations and train for a total of 1,000 epochs (50,000 iterations). The learning rate is reduced each epoch according to the polynomial learning rate (polyLR) schedule presented in ((Chen *et al.*, 2018), see also see Supporting Information S3.1).

### 3.1 Data augmentation

Due to the limited number of training cases extensive data augmentation is performed on the fly during training using the 'batch generators' framework (Isensee *et al.*, 2020). Specifically, we use random rotations, scaling, elastic deformation, gamma augmentation, Gaussian Noise, Gaussian blur, mirroring, random shift and cutout (DeVries & Taylor, 2017) (for details regarding the data augmentation pipeline, see Supporting Information Section S4).

### 3.2 Inference

For model development we used stratified five-fold cross-validation on the training set. The resulting five models are used as an ensemble for test set predictions. We furthermore use test-time data augmentation (mirroring). Ensembling is implemented via softmax averaging, followed by thresholding at 0.5 to obtain the final predictions (see Supporting Information Sections S3.2 and S3.3).

## 4. CNN variant: identifying more single hits

The CNN model described above is optimized for maximizing the F1-score on our training cross-validation. We subsequently refer to it as "MaxF1". In addition, we trained a second CNN model that predicts a larger number of single hits ("moreSH") and leans more towards higher recall values. To achieve that, we made modifications to the sampling strategy as well as the loss function. Specifically, we increased the probability of selecting single hits when constructing the minibatches from 2% to 5% and make use of a weighted cross-entropy loss which weights samples of ground truth single hits higher during loss computation (weights 0.1 and 0.9 for non-single hits and single hits, respectively). For both models (maxF1 and moreSH) we used the same augmentation and inference scheme.

## 5. Comparison metrics of different data selections

To additionally compare different data selections, we also looked at the intersection over union metric, which can be described as:

$$IoU = \frac{Intersection}{Union}, \qquad (4)$$

where *Intersection* is the number of diffraction patterns in the intersection of data sets, and *Union* is the number of diffraction patterns in the union of data sets.

As a result of single hits classification, we obtained data selections with different number of diffraction patterns. In order to compare these selections, we plotted and analyzed the power spectral density (PSD) function, i.e. the angular averaged intensity. To quantify the contrast values of the PSD functions for each selection we introduced the following metrics, which describes the mean difference between the local minimas and maximas over the first three pairs

$$\gamma = \frac{1}{N}\sum_{i=1}^{N}\frac{I_{max}-I_{min}}{I_{max}+I_{min}}, \qquad (5)$$

where N=3 is the number of pairs, $I_{max}$ and $I_{min}$ are values of the PSD function in maximas and minimas, respectively. By looking at PSD functions and corresponding contrast values we can compare various single hits selections and analyse which one has more features.

## 6. Particle size determination

The particle size filtering is also an important part of the SPI data analysis workflow (see Fig. 1 and Supporting Information Section S4). It can help to remove unnecessary diffraction patterns corresponding to other particles apart from viruses under investigation. Previous approach (Fig. 1, black arrows) implied single hit diffraction patterns classification on a certain size range of viruses from 55 nm to 84 nm (see (Assalauova *et al.*, 2020)) and particle size determination was performed on the whole data set *D*. In this work we used the CNN classification after the initial preprocessing step and particle size filtering was applied afterward. Here we used the same results for the virus size estimation as in (Assalauova *et al.*, 2020) and the same virus size range (55 - 84 nm) was considered here.

# IV. Results

## 1. CNN performance

Table 2 summarizes the performance of our CNNs on the training set cross-validation. The MaxF1 configuration obtains balanced precision and recall and a $F_1$-score of 0.645. The number of predicted single hits (120) is close to the number of single hits (100) in this data set. The moreSH configuration, however, trades a higher recall with lower precision, resulting in an overall decreased *F1*-score of 0.536. As expected, the number of predicted single hits is higher being 221 in this case.

**Table 2.** Five-fold cross-validation results (*N*=20,000 training samples).

|  | **MaxF1** | **moreSH** |
|---|---|---|
| **F1-score** | 0.645±0.074 | 0.536±0.018 |
| **Precision** | 0.591± 0.062 | 0.391±0.023 |
| **Recall** | 0.710± 0.096 | 0.860±0.065 |
| **Predicted single hits** | 120 | 221 |

Test set predictions were obtained by ensembling the five models obtained during cross-validation (see Section 3.2). On the test set (171,183) the MaxF1 configuration obtained an F1-score of 0.731 with balanced precision and recall (Table 3). Interestingly, the F1-score is substantially higher than on the training set cross-validation which we attribute to the use of ensembling. The predicted number of single hits (1,257) is close to the number of reference single hits (1,293).

The moreSH configuration expectedly again displays an imbalance between precision and recall. Overall its recall is higher (0.841 vs 0.721) but its *F1*-score is lower at 0.644 (vs 0.731). Again, as expected, the number of predicted single hits is larger (2,086 patterns).

On a workstation equipped with an AMD Ryzen 5800X CPU, 32GB of RAM and a Nvidia RTX 3090 GPU, training each individual model took less than 25 minutes (<2.5h for all 5 models in the cross-validation). Inference speed was ~450 diffraction patterns/s for the ensemble and with test time data augmentation (5 models and mirroring along all axes for a total of 20 predictions per pattern). Predicting the 171k test patterns took less than 7 minutes. Single model prediction without test time augmentation is ~8700 patterns/s and could be considered in case faster inference is required. We should note that training merely required 3.5GB of VRAM and much smaller GPU would have been sufficient as well.

**Table 3.** Test set results ($N$=171,183 test samples).

|  | **MaxF1** | **moreSH** |
|---|---|---|
| **F1-score** | 0.731 | 0.644 |
| **Precision** | 0.741 | 0.522 |
| **Recall** | 0.721 | 0.841 |
| **Predicted single hits** | 1,257 | 2,086 |

## 2. PSD comparison, EM and particle size filtering

As a result of CNN classification, we obtained two data sets: MaxF1 and moreSH with the number of single hit diffraction patterns 1,257 and 2,086, respectively (see Table 4). Plotted PSD functions for both selections are shown in Fig. 4 (blue dashed lines). Additionally, we plotted PSD function for the $D_M$ and $D_{EM}$ selection (Assalauova *et al.*, 2020) containing 1,393 and 1,085 diffraction patterns, respectively (Fig. 4, purple and brown solid lines). The corresponding contrast values and number of patterns for all three data sets (MaxF1, moreSH, $D_M$, $D_{EM}$ selection) are given in Table 4. From Fig. 4 we observe the same number of fringes as in our previous paper, however, the contrast values were lower in the case of CNN classification in comparison to EM classification. As it is expected, PSD functions for MaxF1 and moreSH mimic the behaviour of PSD function of $D_M$ selection which was used as the ground truth for CNN training.

In order to increase contrast in PSD functions from the CNN selection we applied EM-algorithm to MaxF1 and moreSH data sets, respectively (see Supporting Information Section S5), and results of this additional selection are summarized in Fig. 4 (green dashed lines) and Table 4 with notation "+ EM". As we can observe, the contrast $\gamma$ for moreSH plus EM selection showed a substantial improvement (0.64 vs 0.59 without EM) and we also observed a slight improvement for the MaxF1 plus EM selection (0.64 vs 0.63 without EM). At the same time, the EM selection (Assalauova *et al.*, 2020) still has the best result in terms of contrast $\gamma$.

At this point it has to be noted that EM classification in (Assalauova *et al.*, 2020) was performed on a certain size range of viruses from 55 nm to 84 nm that was determined prior to EM classification. To perform particle size analysis in this work, we first plotted histograms of the particle size distribution for each data set (MaxF1 with/without EM algorithm applied, moreSH with/without EM algorithm applied) in Fig. 5. It can be seen that each data selection consists of diffraction patterns within a wide size range. This means that even after single hit classification (with/without EM algorithm), data sets contain diffraction patterns that correspond to particles of different sizes. To be consistent with our previous work the range of the sizes from 55 nm to 84 nm was considered for further analysis and particle size selection was applied. The corresponding PSD functions are plotted in Fig. 4 (solid orange and red lines) and the resulting number of diffraction patterns and contrast values are summarised in Table 5 with notation "+ size selection".

From Fig. 4(a) and from Table 4 it can be seen that for the MaxF1 data set the particle size filtering practically did not changed the contrast values ($\gamma = 0.64$). However, for the selection moreSH with EM algorithm applied the particle size filtering showed the best PSD contrast value $\gamma = 0.65$.

Even though we were able to increase the PSD contrast through different classification strategies and particle size filtering, we, unfortunately, reduced the number of diffraction patterns along the way. For the MaxF1 data set we started from the data set of 1,257 patterns and, finally came to 827 patterns. For moreSH selection we started with the number of patterns of 2,086 and finally came to 1,090 patterns. In the context of our data processing pipeline, where a large number of single hits is required to get reliable results, this can be detrimental.

In the following, we will consider four final data sets: MaxF1 with size filtering applied (Fig. 4(a), orange solid line; Fig. 5(a), orange histogram), MaxF1 with EM algorithm and size filtering applied (Fig. 4(a), red solid line; Fig. 5(a), red histogram), moreSH with size filtering applied (Fig. 4(b), orange solid line; Fig. 5(b), orange histogram), and moreSH with EM algorithm and size filtering applied (Fig. 4(b), red solid line; Fig. 5(b), red histogram).

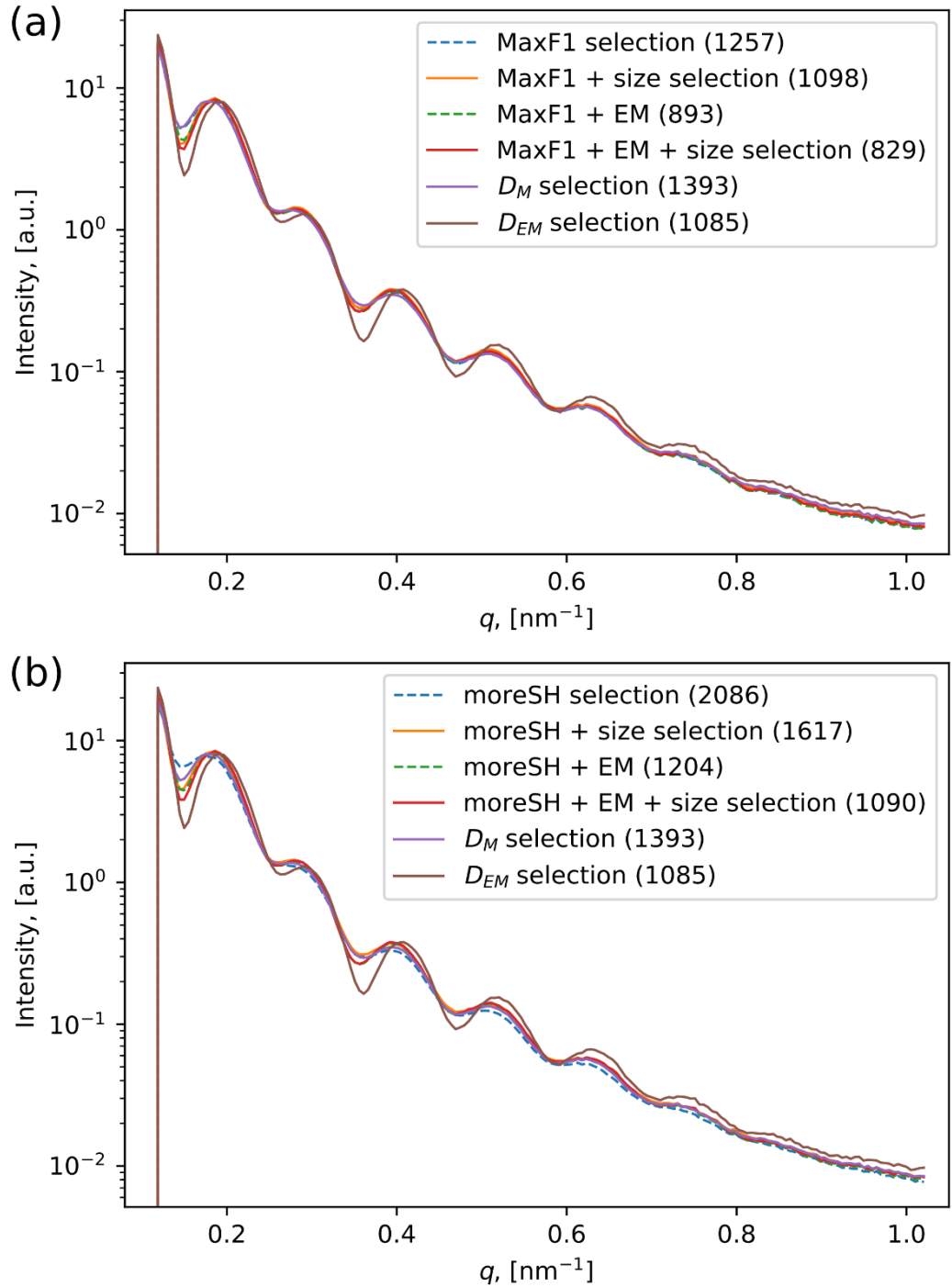

**Figure 4.** PSD functions for the different data sets. (a) PSD functions for the MaxF1 data selection shown in the same way (blue dashed line - the whole selection, orange line - selection with the size filtering applied, green dashed line - selection with the EM algorithm applied, red line - selection with the EM algorithm and size filtering applied). (b) PSD functions for the moreSH data selection (blue dashed line - the whole selection, orange line - selection with the size filtering applied, green dashed line - selection with the EM algorithm applied, red line - selection with the EM algorithm and size filtering applied). Both (a) and (b) panels contain the PSD function of the $D_M$ and $D_{EM}$ selections. In the legend numbers of diffraction patterns for each selection are shown in brackets.

**Table 4.** Number of diffraction patterns in different data sets of single hits and PSD contrast values for each of them.

| Data set | Number of diffraction patterns | PSD contrast $\gamma$ |
|---|---|---|
| MaxF1 | 1,257 | 0.63 |
| MaxF1 + EM | 893 | 0.64 |
| MaxF1 + size selection | 1,098 | 0.64 |
| MaxF1 + EM + size selection | 829 | 0.64 |
| moreSH | 2,086 | 0.59 |
| moreSH + EM | 1,204 | 0.64 |
| moreSH + size selection | 1,617 | 0.62 |
| moreSH + EM + size selection | 1,090 | 0.65 |
| $D_M$ | 1,393 | 0.59 |
| $D_{EM}$ | 1,085 | 0.71 |

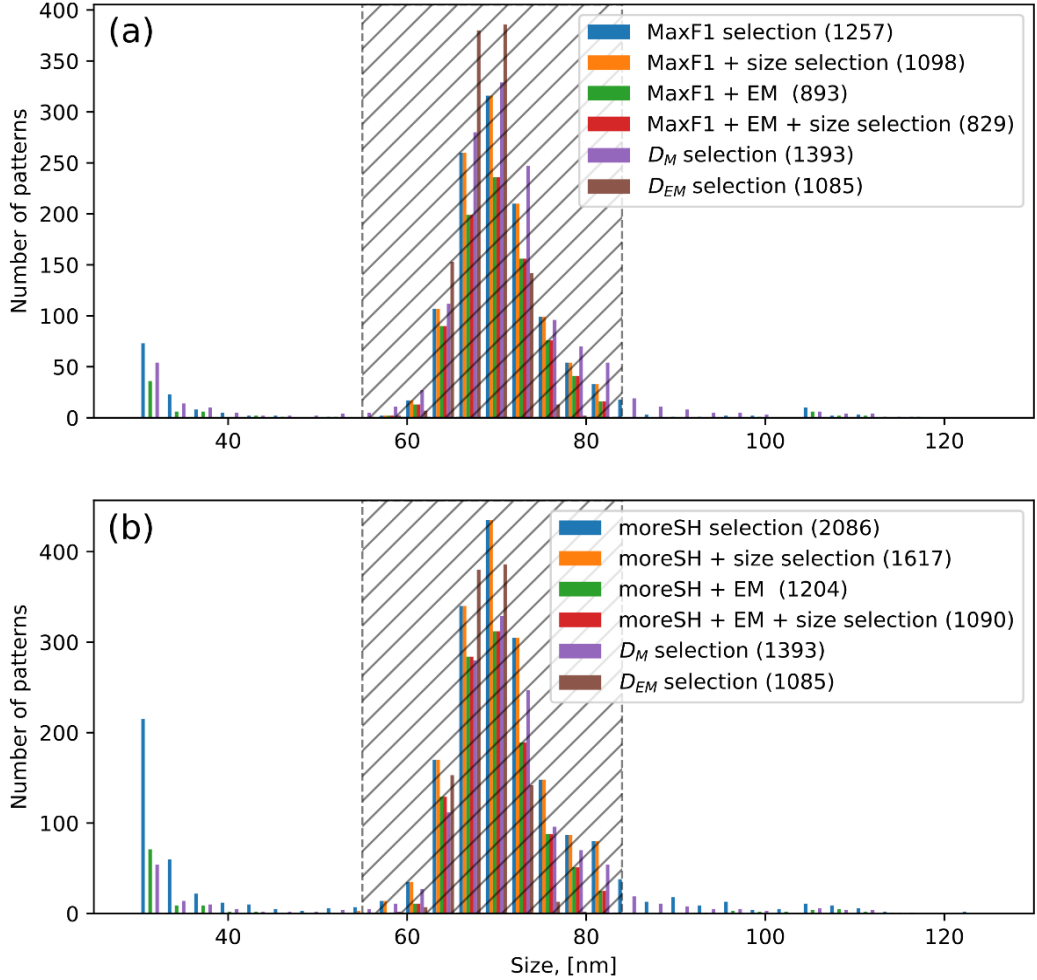

**Figure 5.** Particle size histograms for different data sets. (a) Particle size histogram for the MaxF1 data selection shown in the same way (blue - the whole selection, orange - selection with the size filtering applied, green - selection with the EM algorithm applied, red - selection with the EM algorithm and size filtering applied). (b) Particle size histogram for the moreSH data selection (blue - the whole selection, orange - selection with the size filtering applied, green - selection with the EM algorithm applied, red - selection with the EM algorithm and size filtering applied). In both (a) and (b) panels dashed areas indicate the particle size range from 55 nm to 84 nm; $D_M$ selection is shown in purple bins; $D_{EM}$ selection is shown in brown bins. In the legend, the number of diffraction patterns for each selection is given in brackets.

## 3. Intersection over union comparison

We also compared diffraction patterns in our four final data sets in terms of intersection over union metric. Obtained for different pairs of data sets it is shown in Table 5. We also calculated the intersection over union over three selections: MaxF1 with size filtering applied, moreSH with size filtering applied and $D_{EM}$ selection that gave to us IoU=29% with 575 diffraction patterns in the intersection. For another three selections: MaxF1 with EM algorithm and size filtering applied, moreSH with EM algorithm and size filtering applied and $D_{EM}$ selection, gave to us IoU=29% with 469 diffraction patterns. We think that this choice of diffraction patterns in the intersection of three data selections is providing us with the most important diffraction patterns that contain the features of virus structure from all data selections.

**Table 5.** Number of diffraction patterns in intersections of different pairs of data sets. Initial number of diffraction patterns in the sets is shown in the brackets. In the second line IoU is shown.

|  | MaxF1 + size selection (1,098) | MaxF1 + EM + size selection (829) | moreSH + size selection (1,617) | moreSH + EM + size selection (1,090) | $D_M$ selection (1,393) | $D_{EM}$ selection (1,085) |
|---|---|---|---|---|---|---|
| **MaxF1 + size selection (1,098)** | 1,098 100% | 829 75% | 1,097 68% | 878 67% | 875 54% | 575 36% |
| **MaxF1 + EM + size selection (829)** | 829 75% | 829 100% | 829 51% | 730 61% | 678 44% | 485 34% |
| **moreSH + size selection (1,617)** | 1,097 68% | 829 51% | 1,617 100% | 1,090 67% | 1,006 50% | 686 34% |
| **moreSH + EM + size selection (1,090)** | 878 67% | 730 61% | 1,090 67% | 1,090 100% | 791 47% | 651 43% |
| **$D_M$ selection (1,393)** | 875 54% | 678 44% | 1,006 50% | 791 47% | 1,393 100% | 574 30% |
| **$D_{EM}$ selection (1,085)** | 575 36% | 485 34% | 686 34% | 651 43% | 574 30% | 1,085 100% |

## 4. Orientation determination

The next step of the workflow for the SPI analysis after the single hit classification is orientation determination of the diffraction patterns (see Fig. 1). In SPI experiments particles are injected into the X-ray beam in random orientations, so to retrieve a 3D intensity map of the virus from the selected 2D diffraction patterns, orientation recovery has to be done. Expand-maximize-compress (EMC) algorithm (Loh & Elser, 2009) in the software Dragonfly (Ayyer *et al.*, 2016) was used to retrieve the orientation of each diffraction pattern and to combine them into one 3D intensity distribution of the PR772 virus. We retrieved the orientation of all previously selected data sets with the size filtering applied, with and without the EM classification applied.

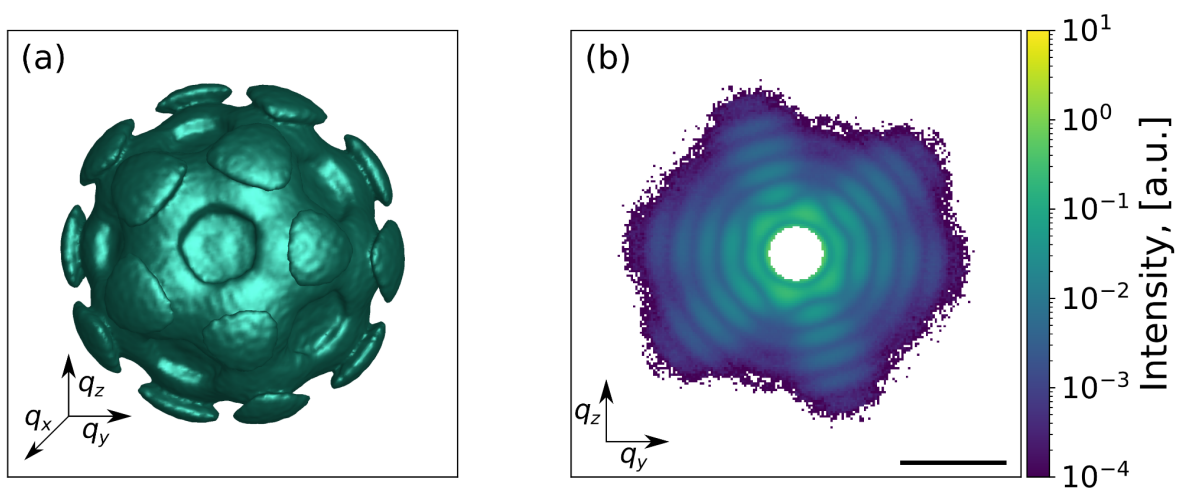

**Figure 6.** Reciprocal space representation for the MaxF1 selection with the EM algorithm and size filtering applied. (a) 3D intensity distribution in reciprocal space of the virus after background subtraction, (b) 2D cut of the distribution. All diffraction patterns are shown in logarithmic scale. Black scale bar in (a) and (b) denotes 0.5 nm$^{-1}$.

Visual inspection does not allow us to see a significant difference between data sets (MaxF1 and moreSH with/without EM algorithm applied, and with size filtering applied). However, for all four data sets the background at high q-values is well seen (see Supporting Information Fig. S4). Background subtraction is a common task in SPI data analysis and several techniques were already developed (Rose *et al.*, 2018; Lundholm *et al.*, 2018; Ayyer *et al.*, 2019). In this work we defined the level of the background as the mean signal in the high q-region, where the presence of meaningful signal from the particle is negligible. The orientation determination results after background subtraction on the MaxF1 CNN selection with the EM and size filtering applied is shown in Fig. 6 (see for other data sets Supporting Information Figure S5).

## 5. Phase retrieval and reconstructions

The next and the final step in our workflow is the phase retrieval and reconstruction of electron density of our virus particle from the 3D reciprocal space data (see Fig. 1). Since the experimental measurements provide only the amplitude of the complex valued scattered wave field we applied iterative phase retrieval algorithms (Fienup, 1982; Marchesini, 2007) in order to determine the 3D structure of the virus particle. The following algorithms were used in this work for the phase retrieval: continuous hybrid input-output (Fienup, 2013), error reduction (Fienup, 1982), Richardson-Lucy deconvolution (Clark *et al.*, 2012), and shrink-wrap (Marchesini *et al.*, 2003).

Further, we proceeded in the same way as in the work (Assalauova *et al.*, 2020). The phase retrieval procedure consisted of two steps. In the first step, the central gap in the 3D intensity map of the virus that originated from the masking of the initial 2D diffraction patterns was filled. Running 3D reconstruction with a freely evolving central part produced a signal in the masked region which was used further. In the second step, the 3D intensity maps with the filled central part were used to perform phase retrieval. We first performed 50 reconstructions for each intensity map and then used mode decomposition (Khubbutdinov *et al.*, 2019; Assalauova *et al.*, 2020) to determine the final 3D electron density structure of the virus.

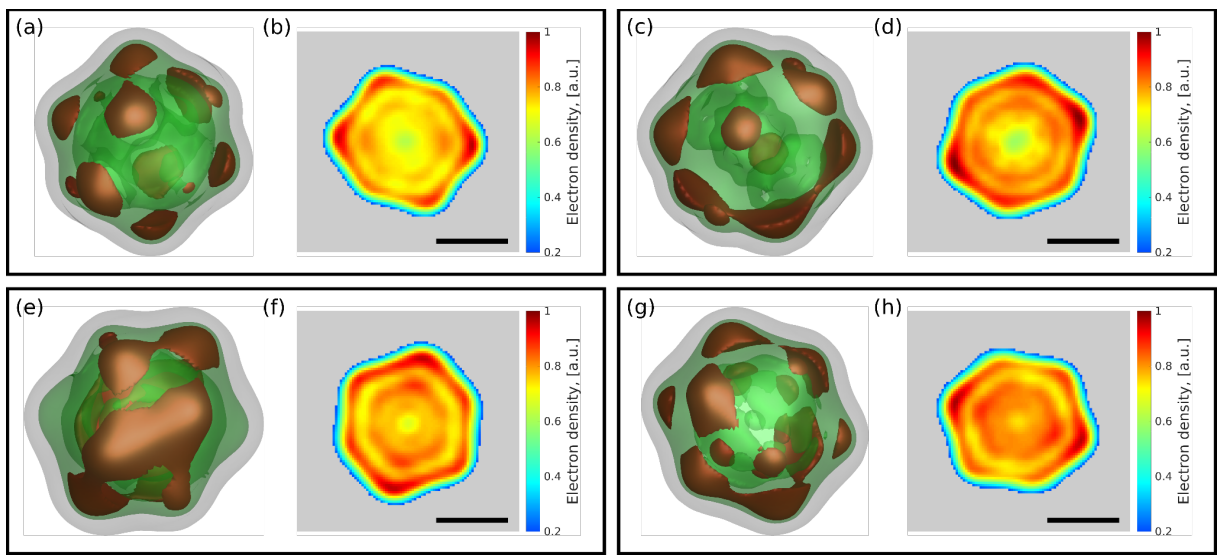

**Figure 7.** PR772 virus reconstructed from the different data sets. (a-d) Reconstruction of single hit diffraction patterns selected by MaxF1 with size filtering applied (a, b) and MaxF1 with EM algorithm and size filtering applied (c, d). (e-h) Reconstruction of the single hit diffraction patterns selected by moreSH with the size filtering applied (e, f) and moreSH with EM algorithm and size filtering applied (g, h). (a, c) The inner structure of the virus with 88% (brown), 75% (green), 20% (grey) levels of intensity for MaxF1 selections. (e, g) The inner structure of the virus with 86% (brown), 75% (green), 20% (grey) levels of intensity for moreSH selections. (b, d, f, h) 2D slices of the corresponding structure with the same scale bar of 30 nm. For visual representation each virus structure was three times upsampled.

The final virus structure for each data selection, obtained in the described way, is shown in Fig. 7. All expected features are present in these reconstructions: icosahedral structure of the virus, higher density in the capsid part of the virus, and reduced density in the central part. Resolution of the obtained images, evaluated by the Fourier-shell correlation (FSC) method, gave the values from 6 nm to 8 nm (see Supporting Information Section S7). The slightly higher resolution determined in this work relative to our previous work (6.9 nm) may be related to the comparably small number of diffraction patterns used in the FSC method. As we observe in Fig. 7(a-d), the electron density of the virus in the CNN MaxF1 selection and MaxF1 with EM selections plus size filtering in both cases are practically identical. We see small differences to the previous electron density in the CNN moreSH selection and moreSH with EM selections plus size filtering in both cases (Fig. 7(e-h)). At the same time, the central slice in all four reconstructions (Fig. 7(b,d,f,h)) is practically the same, showing the same size of the capsid layer. Since we have 400 - 500 diffraction patterns in common with the considered data selections and our previous work (Assalauova *et al.*, 2020), we can assume that these were the ones which contributed and shaped the final reconstructed results in such a common way for all five data selections.

### V. Discussion and summary

Our studies with the CNN-based single hit classification implemented within SPI data analysis workflow resulted in a reasonable structure reconstruction of the virus PR772 (see Fig. 7).

We compared two competing CNN selections, MaxF1 and moreSH. The MaxF1 selection was intended to select single hits with an optimal *F1*-score. The selection moreSH was optimised for finding more single hit diffraction patterns (high recall). Both selections were refined by applying the EM algorithm and limiting the selection to particle sizes in the range 55-84 nm (Table 4). Driven by the need for many single hits in the reconstruction pipeline, the moreSH configuration was conceived with the intention to miss as few single hits as possible and clean up the selection afterwards using EM selection and size filtering, hoping to achieve a higher resolution than its maxF1 counterpart. Unfortunately, this goal was missed: maxF1 yielded approximately the same resolution even though moreSH approach resulted in 1,090 instead of the 829 selected single hits found by maxF1 (after EM and size selection applied). We can therefore conclude that optimising balanced precision and recall through maximising the *F1-score* is a suitable target for model development.

CNNs learn from their given training dataset. Unfortunately, the selection provided by (Li *et al.*, 2020) which was used for this purpose here, as any other manual selection, can be subjective. It could be that the task of identifying single hits is not necessarily identical to the task of finding the ideal set of patterns needed for reconstruction. In an ideal world, the CNNs should be trained with the patterns ideally suited for reconstruction. Until we identify a way of obtaining ideal patterns from a subset of our data, subjectively selected single hits are the next-best solution. We would also like to note that the values of precision and recall summarized in Table 3 are far from the best values achieved in classification of natural images by means of CNN.

The particle size filtering step is quite important and has to be applied throughout the SPI analysis pipeline. The real experiment may run in the following way. A trained person will be selecting a number of single hits as well as non-single hits and then will run the CNN selection on the diffraction patterns coming from the experimental stream. After the size filtering this selection will be uploaded to the SPI workflow as shown in Fig. 1 and final electron density of a single particle will be obtained as a result.

Reconstructing the 3D structure from a selection of single hits is expensive: both computationally as well as in terms of manual labour. We introduced the PSD contrast in the hope it would constitute a good substitute measure for the quality of a selection. If successful, this would have allowed us to optimise our CNNs more directly towards identifying an optimal set of single hits for reconstruction through maximising their PSD contrast. Comparing the PSD contrast between CNN selections and $D_{EM}$ selection (Assalauova *et al.*, 2020) revealed that the contrast in CNN selection is always lower than in EM selection. We initially thought that this may be problematic for the reconstructions. However, as results in Fig. 7 demonstrate, this is not the case and our CNN selection is working well, resulting in an electron density of the PR772 virus that is similar to our previous work (Assalauova *et al.*, 2020). These results indicate that the PSD contrast may not be a good substitute for reconstruction fidelity. Deviations from a circular shape, as are present in PR772 might explain this observation.

We have proposed a SPI workflow that uses a CNN-based single hit classification at an earlier stage of the data analysis pipeline. This approach can be beneficial not only because it can be run during SPI experiments but also because it can significantly reduce the number of diffraction patterns for further processing. That is important for data storage as soon as the size of collected data during one experiment at a megahertz XFEL facility can easily reach several petabytes. Another convenience using CNNs for single hit classification is that the network can be trained on a relatively small amount of data at the beginning of the SPI experiment and can be simply applied throughout the rest of the experiment.

Bringing non-standard AI-based solutions in a well-described SPI analysis workflow may be beneficial for the future development of the SPI experiments. Here we demonstrated the use of CNN at the single hit diffraction patterns classification step which can be applied not only after the experiment but, importantly, also during an experiment and can significantly reduce the size of data storage for further analysis stages. That could be an important advantage with the development of high repetition rate XFELs (Decking et al., 2020) with data collection with the Megahertz rate (Sobolev *et al.*, 2020). Handling experimental data with the CNNs also saves computational time: once the CNN is trained and new data are obtained, there is no need to retrain CNN again as it is needed with other classification approaches.

**Author contributions**

I.A.V. conceived the presented idea. F.I., A.I. and D.T. performed the development of the CNN and S.B. performed EM analysis. F.I., D.A., and A.I. analysed the obtained results. F.I., D.A., A.I., and I.A.V. wrote the paper. All authors read and agreed on the final text of the paper.

**Data availability**

All data are available in (Li *et al.*, 2020). The source code associated with the CNNs will be made available upon publication.

**Conflicts of interest**

The authors declare no conflicts of interest.

**Acknowledgements**

The authors are thankful to E. Weckert for the support of this Project. I. A. V. acknowledges the financial support of the Russian Federation represented by the Ministry of Science and Higher Education of the Russian Federation (Agreement No. 075-15-2021-1352). The authors are thankful to Luca Gelisio for careful reading of the manuscript. Part of this work was funded by Helmholtz Imaging (HI), a platform of the Helmholtz Incubator on Information and Data Science.

Nishiyama, T., Ovcharenko, Y., Piseri, P., Plekan, O., Prince, K. C., Stienkemeier, F., Ueda, K., Callegari, C., Möller, T. & Rupp, D. (2019). *Phys. Rev. E*. **99**, 063309.

# Supporting Information

### S1. Data cropping

Data was prepared before applying CNN-based single hit diffraction patterns classification. The region of interest on all diffraction patterns from the data set D was cropped as it is shown in Fig. S1.

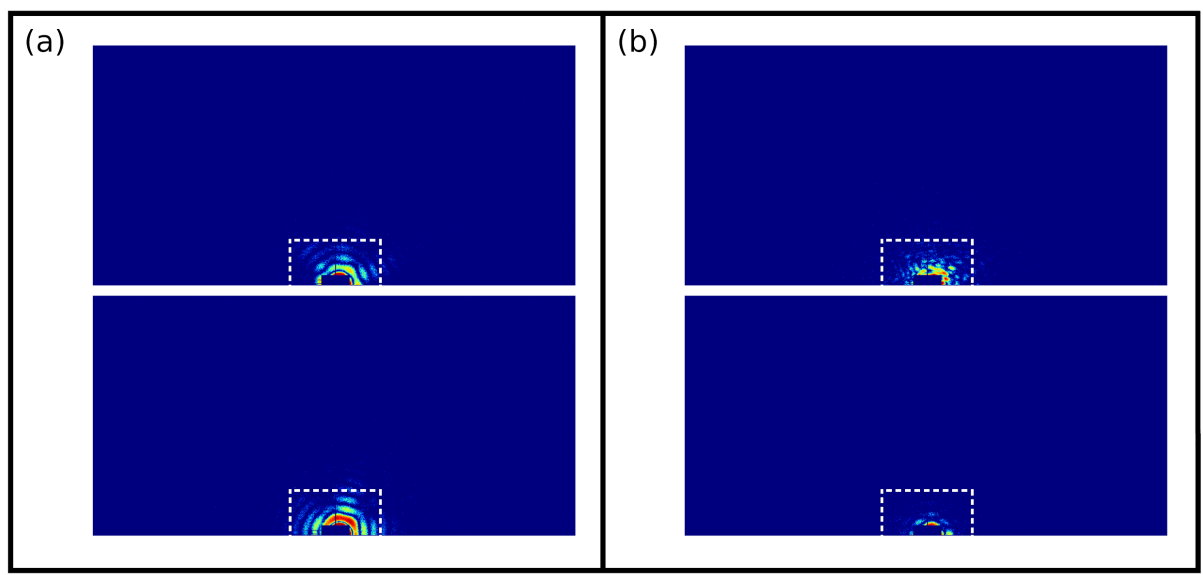

**Figure S1.** Illustration of data cropping before sending to the input of a CNN. The center of diffraction pattern is located around the center of the bottom part of the only operational detector plane with dimensions in pixels 1024 × 512. The area surrounding the center of each diffraction pattern with dimensions in pixels 192 × 96 is cropped (white dotted rectangle). The cropped part is used as an input of a CNN. (a) Single hit examples, (b) non-single hit examples.

### S2. VGG

The main studies in the field of CNN classification of single hits were carried out with the network architecture pre-activated ResNet-18 described in the main part of the paper. In order to investigate an effect of CNN depth required for the specific task of single hit classification, a VGG-style network was implemented within the same pipeline. This network is realized as a plain sequence of convolutional layers organized in four downsampling stages (Fig. S2). The activation function is ReLU. Batch normalization layer precedes each convolutional layer, except the first one. Dimensionality reduction is realized via maximal pooling. The number of filters in the convolutional layers of the first stage is 16. It rapidly grows up to 256 at the last stage. This growth is intentionally fast. It allows to extract more higher level features while preventing the network from growing in depth. Global average pooling is used to linearize the final feature representation of the shape 12 x 16 to a feature vector used for classification.

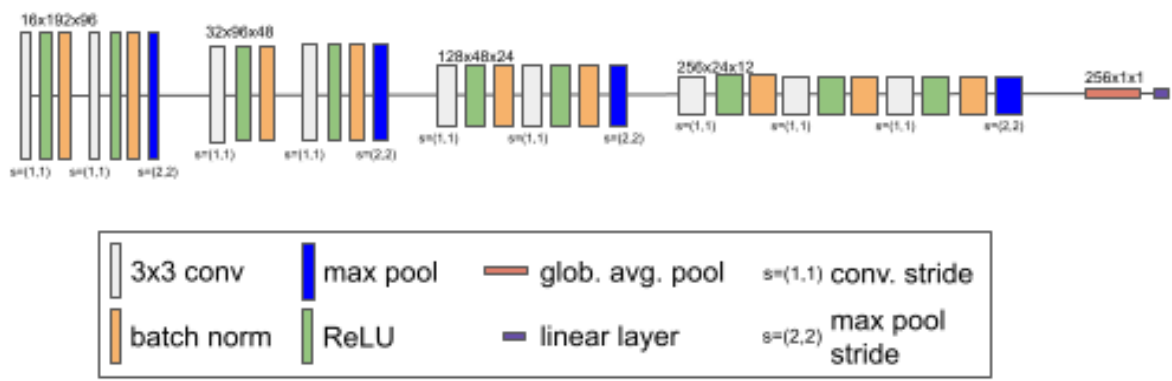

**Figure S2.** VGG-style network architecture. We use a simple VGG-style network for comparison. It has the same input size of 192 x 96. It processes the input in four downsampling stages. Downsampling is implemented via maximal pooling. The convolutional layer of the first stage has 16 filters. The number grows up to 256 filters for the fourth stage. Global average pooling is used to linearize the final feature representation of the shape 12 x 16 to a feature vector used for classification.

Training, validation and test follow the same procedure described for Resnet-18. The results for relevant metrics for five-fold cross-validation is shown in Table S1. Test performance metrics for the VGG-style network (Table S2) is similar to that of Resnet-18 (Table 3 in the main text). This is an indication that the choice of network depth within the investigated limit has negligible effect. Thus, a simple VGG-style network can be sufficient for the task.

**Table S1.** VGG five-fold cross-validation results (n=20,000 training samples).

| | |
|---|---|
| *F1*-score | 0.72 |
| Precision | 0.656 |
| Recall | 0.678 |
| Predicted single hits | 113 |

**Table S2.** VGG test set results (n=171,183 test samples).

| *F1*-score | 0.727 |
|---|---|
| Precision | 0.78 |
| Recall | 0.681 |
| Predicted single hits | 1,130 |

### S3. Details on CNN training and validation

We train the network with stochastic gradient descent using the Adam optimizer (Kingma & Ba, 2014), a minibatch size of 64 and an initial learning rate of $10^{-4}$. The standard cross-entropy loss function is used. Samples within minibatches are sampled randomly with replacement. We modify the sampling probabilities such that on average 2% of the presented samples are single hits. We define an epoch as 50 training iterations and train for a total of 1,000 epochs (50,000 iterations). The learning rate is reduced each epoch according to the polyLR schedule presented in (Chen *et al.*, 2018).

For model development we used five-fold cross-validation on the training set. The resulting five models are then used as an ensemble for test set predictions. Ensembling is implemented via softmax averaging, followed by thresholding at 0.5 to obtain the final predictions.

Below are the details on learning rate scheduler, five-fold cross-validation procedure and softmax averaging implemented to make the final prediction of the CNN model.

### S3.1. Polynomial learning rate (polyLR) policy

Learning rate is one of the most important hyper-parameters in any neural network optimization process. It controls the speed of network convergence in the training process. One of the most common algorithms of minimization of loss function is stochastic gradient descent (SGD). SGD first computes the gradients of the loss function with respect to all model parameters using an algorithm called back-propagation and then updates the model weights w as follows:

$$w^{i+1} = w^i - \eta \cdot \frac{\partial L}{\partial w}, \quad (S1)$$

where L is the loss function, i is iteration number, nu is learning rate. A conventional approach to control convergence of a model is to set an initial value of learning rate and let it decrease over time. Here we use a learning rate scheduler called polynomial learning rate policy (polyLR) (Chen *et al.*, 2018). The learning rate is changed during training according to the equation:

$$\eta = \eta_0 \cdot (1 - \frac{i}{T_i})^{power}, \qquad (S2)$$

where $T_i$ is the total number of iterations during training.

**S3.2. K-fold cross validation**

Cross-validation is a procedure used to evaluate machine learning models on a limited dataset size, i. e. the amount of data is too small to draw robust conclusions using a conventional training and validation split. The procedure of k-fold cross-validation is the following. The entire data set available for training and validation is shuffled and split into k groups. For each unique group, the data from this group becomes the validation data set; the respective training data set consists of the other k-1 groups. As a result, there are k individual trained models. The performance metrics are then defined by average performance of these models.

We chose k=5 for developing our models. Final performance on the test set is obtained by using the resulting five models as an ensemble, as described in the following section.

**S3.3. Ensembling via softmax averaging**

Ensembling refers to combining predictions from multiple machine learning models. It is a commonly used strategy to reduce the variance of the models and increase the overall quality of the predictions. In the case of image classification (which is the setting used in this publication), ensembling can be implemented via softmax averaging. Here, this is implemented in the following way: Each CNN model issues a prediction for each diffraction pattern of the test data set providing single hit probability (ranging from 0 to 1). The average of five predictions, one for each model, is then the single hit probability for the ensemble. We put a threshold for the average single hit probability to obtain the final prediction. Diffraction patterns with final probability above 0.5 are classified as single hits.

### S3.4. Details on data augmentation

Data augmentation is a powerful tool to improve the robustness of models trained on a limited number of training cases. By running transformations on the training cases, new images are generated that direct the models to learn better generalizing features and thus ultimately improve their generalization capabilities on the test set. We use the following transformations from the 'batchgenerators' framework[1]: random rotations, scaling, elastic deformation, gamma augmentation, Gaussian noise, Gaussian blur, mirroring, random shift and cutout.

Random rotation is a common augmentation technique when a source image is rotated clockwise or counterclockwise by some number of degrees. This changes the position of the object in the image. In random rotation of the image its corners are cut off, after rotation the new corners are filled with padding.

Scaling can be done outward or inward. When scaling inward, the resultant image size is larger than the original image size. A section is cut out from the resultant image to make the size equal to the original image. When scaling outward, the size of the image is reduced, the missing part is filled with padding.

Obtaining an augmented image using elastic deformations is done in two parts. First, a random stress field is generated for horizontal and vertical directions with randomly sampled values:

$$\Delta_x = G(\sigma) \cdot (\alpha \cdot \text{Rand}(n, m))$$

$$\Delta_y = G(\sigma) \cdot (\alpha \cdot \text{Rand}(n, m)), \tag{S3}$$

where $G(\sigma)$ is the strength of the smoothing operation given by the standard deviation of the Gaussian filter $\sigma$, $\alpha$ is a parameter defining the maximum value for the random initial displacement, n and m are the image dimensions. After that, the stress field is applied to the image by moving each pixel to a new position (Eq. S4) using spline interpolation of order one to obtain pixel values at integer coordinates:

$$I_{deformed}(j + \Delta_x(j,k), k + \Delta_y(j,k)) = I(j,k), \tag{S4}$$

---

[1] https://github.com/MIC-DKFZ/batchgenerators

where *I* and *I*<sub>deformed</sub> are the initial and deformed images, j and k are pixel coordinates.

Gamma augmentation is a nonlinear operation used to encode and decode luminance in images, it is defined by power-law expression:

$$V = AU^{\gamma}, \tag{S5}$$

where V is resultant pixel value, U is initial pixel value, A is a constant, γ is a parameter.

Gaussian noise is an additive noise type, where the intensity value in a pixel with the coordinates (x,y) for the noisy image is given by the expression:

$$N(x,y) = A(x,y) + B(x,y), \tag{S6}$$

where the A(x,y) in the pixel value of the original image, B(x,y) is the added noise. The added value of noise is defined by probability density function of Gaussian random value is indicated in equation:

$$p(z) = \frac{1}{\sqrt{2 \cdot \pi} \cdot \sigma} \cdot e^{\frac{-(z-\mu)^2}{2 \cdot \sigma^2}}, \tag{S7}$$

where σ and μ are standard deviation and mean values, z is pixel value.

Gaussian blur is a type of image-blurring filter that uses a Gaussian function for calculating the transformation to apply to each pixel in the image. The response of the Gaussian filter in two dimensions is described by

$$g(x,y) = \frac{1}{2 \cdot \pi \cdot \sigma^2} \cdot e^{-\frac{x^2+y^2}{2 \cdot \sigma^2}}, \tag{S8}$$

x and y are the distances from the filter origin in the horizontal and vertical directions, respectively. σ is the standard deviation of the Gaussian distribution.

Mirroring implies flipping images along vertical and horizontal axes.

Random shift is a transformation when the image as a whole is shifted horizontally and vertically by a random number of pixels. The missing parts at the edges appeared due to these shifts being filled with padding.

Cutout is a data augmentation technique that randomly masks out square regions in images. These regions are of random size, appear in random positions in the image and filled with padding.

## S4. Particle size determination

Particle size determination was implemented to the initial set *D* by fitting the power spectral density (PSD) function of each diffraction pattern with the PSD of diffraction pattern from the spherical particles in a range of sizes from 30 to 300 nm and was described in (Assalauova *et al.*, 2020).

## S5. Application of EM algorithm

Task of a single hit classification in cryogenic electron microscopy (cryo-EM) is commonly solved with this approach (Dempster *et al.*, 1977; Scheres *et al.*, 2005). The EM classification algorithm is designed to distribute the whole data set into a predefined number of clusters. On each iteration, probabilities of patterns to be assigned to each cluster are calculated and cluster models are updated by weighted averaging of the associated patterns, where weights are determined by obtained probabilities. After the algorithm converges, one can manually select the required clusters which correspond to the particle under investigation.

If one considers the contrast of the PSD function as a criterion for best reconstruction, the EM algorithm outperforms CNN classification. EM-based algorithm was applied to the diffraction patterns selected by CNN: MaxF1 and moreSH data sets, containing 1,257 and 2,086 patterns respectively. Both selections were distributed into 20 classes (example of distribution for MaxF1 data set is in Fig. S3) and after 10 iterations of the algorithm, the obtained classes were inspected. Some of them clearly contained diffraction patterns of the virus and the rest ones contained other scattering. Classes of interest were selected manually by the 6-fold symmetry expected from the virus. In the case of the MaxF1 data set, classes 3, 6, 9, 12, 16, 19 (highlighted with red title) were considered to contain patterns of interest. The final numbers of the diffraction patterns before and after applying EM-based algorithm are presented in the main text in Table 4.

Below are computing times to obtain $D_{EM}$ selection by size filtering of 191k diffraction patterns and performing the EM algorithm on 18k patterns in the size range 55-84 nm. Size estimation takes 16 min 26s. It is single threaded and do not really benefit from many cores. Extraction and saving of filtered data take: 20 min 37 s. It is limited by storage read and write speed. EM classification takes 26 min 16 s for 10 iterations. For 5 classifications it is 2 h 11 min 20 s. Calculations were performed on a computer cluster node (max-exfl027) with 2 Intel E5-2698 v4 @ 2.20GHz. It is 40 cores and 80 threads total. The node also has 512GB of memory, but it is barely used by EM.

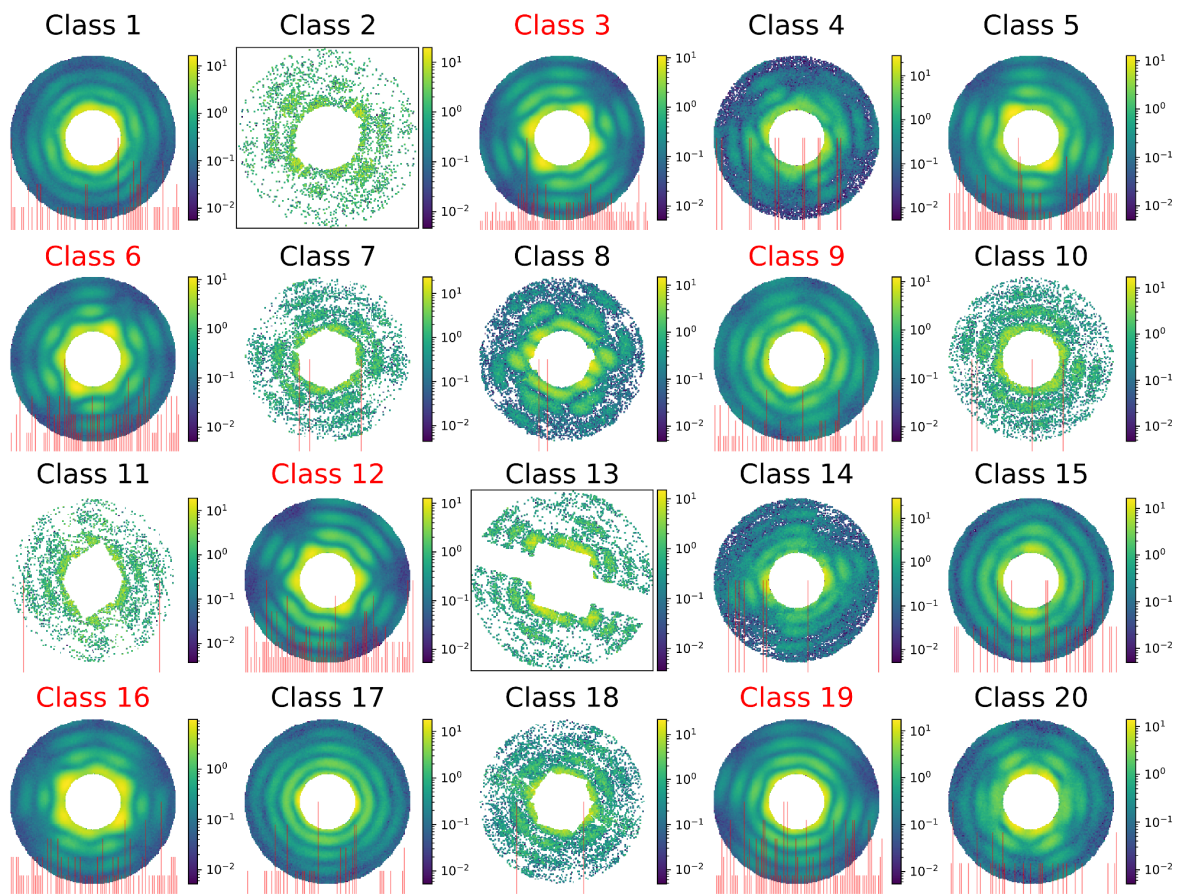

**Figure S3.** EM-based classification of single hit diffraction patterns for MaxF1 data set. Data were distributed into 20 classes, Classes 3, 6, 9, 12, 16, 19 were selected as containing diffraction patterns of PR772. These classes contain 893 patterns in total.

## S6. Orientation determination results for different data selections

Orientation determination of the diffraction patterns was performed by using Expand-maximize-compress (EMC) algorithm (Loh & Elser, 2009) in the software Dragonfly (Ayyer *et al.*, 2016). The result of such a procedure is the 3D intensity distribution of the investigated particle (Fig. S4).

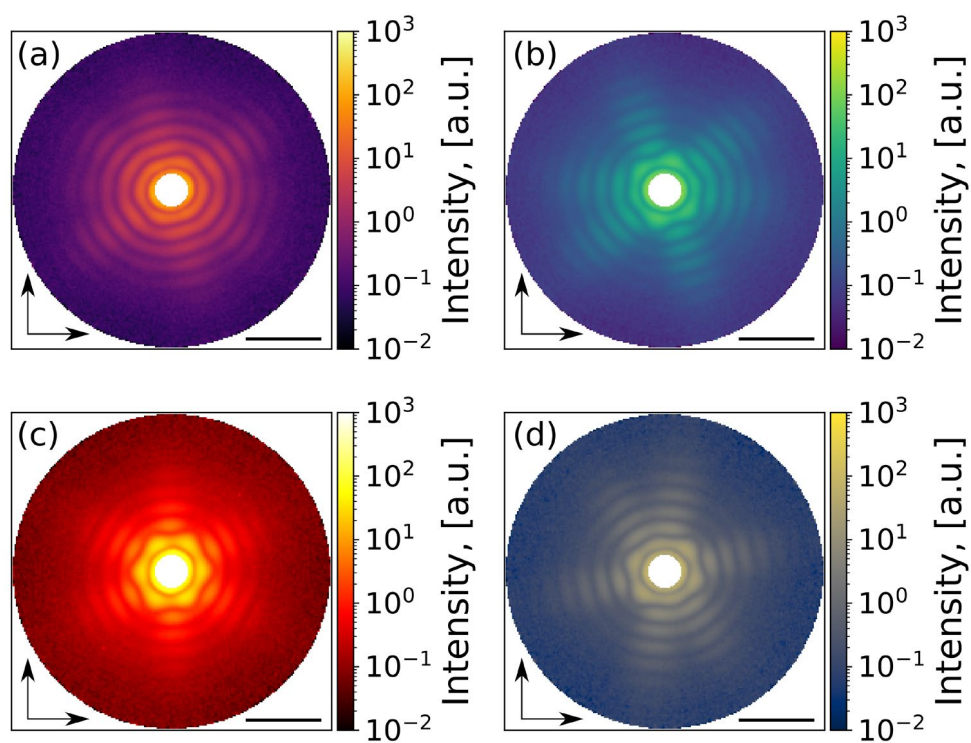

**Figure S4.** Result of orientation determination. 2D central slice of 3D intensity distribution for MaxF1 with the size filtering applied (a), MaxF1 with the EM algorithm and size filtering applied (b), moreSH with the size filtering applied (c), moreSH with the EM algorithm and size filtering applied (d). Black scale bar in denotes 0.5 nm$^{-1}$, vertical and horizontal axes denotes $q_z$ and $q_y$ directions, respectively.

Background level as the mean signal in the high q-region, was subtracted from each data set: MaxF1 with the size filtering applied, MaxF1 with the EM algorithm and size filtering applied, moreSH with the size filtering applied, moreSH with the EM algorithm and size filtering applied; results are shown in Fig. 6 and Fig. S5.

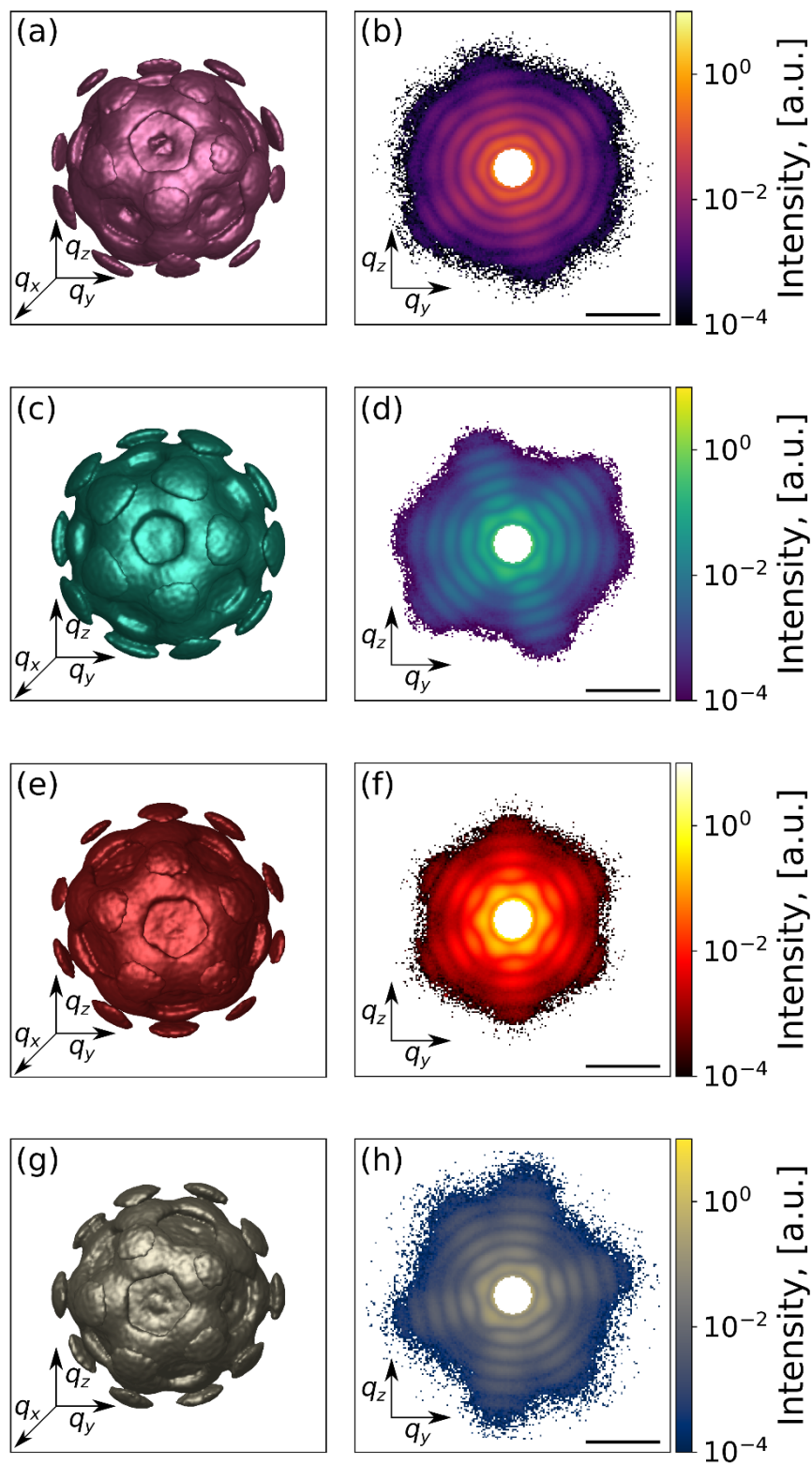

**Figure S5.** Reciprocal space representation for different data selections. (a,b) 3D intensity distribution and its 2D cut of MaxF1 with the size filtering applied data selection. (c,d) 3D intensity distribution and its 2D cut of moreSH with the size filtering applied. (e,f) 3D intensity

distribution and its 2D cut of moreSH with the EM algorithm and size filtering applied. All diffraction patterns are shown in logarithmic scale. Black scale bar in (a,d,f) denotes 0.5 nm$^{-1}$.

### S7. Resolution

In order to numerically measure the difference between electron density reconstruction results from different data selections, we calculated Fourier-shell correlation (FSC) resolution (Harauz & van Heel, 1986). This method implies a data set divided into two sets, each of them oriented, reconstructed and then compared to each other. As for the resolution criterion, the 1/2-bit threshold (van Heel & Schatz, 2005) was used and is related to the signal-to-noise ratio of the two reconstructions. The resolution estimation is the intersection point of the 1/2-bit threshold with the FSC-curve.

Obtained FSC resolution for all four data sets (MaxF1 and moreSH with/without EM algorithm applied, with/without size filtering applied) fluctuates from 5.8 nm to 8 nm and is shown in Table S3 and Fig. S6. Applied EM algorithms for the CNN-based classification could improve the reconstruction result by several nanometers in terms of FSC resolution. And CNN-based single hit diffraction patterns classification by itself with size filtering applied could give quite good resolution. In comparison with the previous EM selection (Assalauova *et al.*, 2020) with 6.9 nm resolution, the results obtained in this work showed overall agreement in virus structure (Fig. 7) and FSC resolution, the difference varies +/- 1 nm. The best result appeared to be MaxF1 with the EM algorithm and the size filtering applied selection - with the FSC resolution of 5.8 nm. Corresponding inner structure (Fig. 7(c)) and 2D central slice (Fig. 7(d)) demonstrated only slight variance from the previous work (Assalauova *et al.*, 2020).

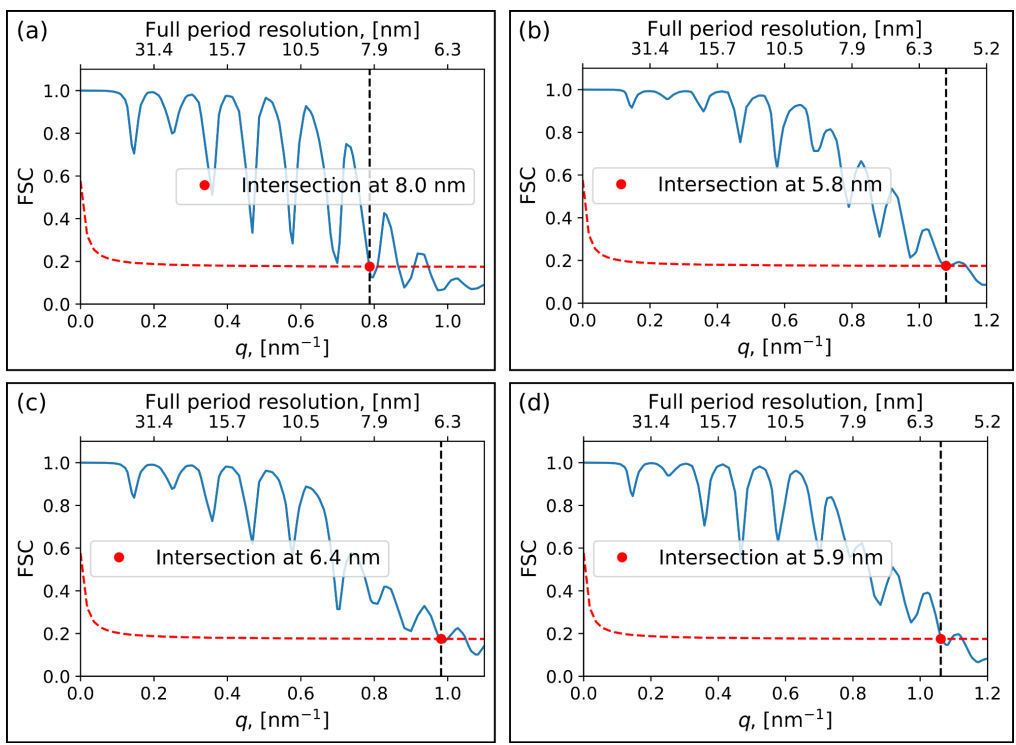

**Figure S6.** Fourier shell correlation resolution for different data selections. In all cases 1/2-bit threshold (red dashed line) was used. (a, b) Resolution for MaxF1 with size filtering applied (a) and MaxF1 with EM algorithm and size filtering applied (b). (c, d) Resolution for moreSH with size filtering applied (c) and moreSH with EM algorithm and size filtering applied (d).

Table S3. FSC resolution for different data selections.

| Data set | FSC resolution, nm |
|---|---|
| MaxF1 + size selection | 8 |
| MaxF1 + EM + size selection | 5.8 |
| moreSH + size selection | 6.4 |
| moreSH + EM + size selection | 5.9 |